\documentstyle[twoside,fleqn,espcrc2]{article}
\title{
Monopole condensation in the 
ground state of gauge theories: a disorder parameter.}
\author{L. Del Debbio,
A. Di Giacomo
and
G. Paffuti
\thanks{
Partially supported by MURST and by EC Contract 
CHEX-CT92-0051}
\address{Dipartimento di Fisica Universit\`a di Pisa and
INFN Sezione di Pisa}
}
\begin{document}
\begin{abstract}
We construct a disorder parameter for dual superconductivity of 
the ground state of $U(1)$ gauge theory.
\end{abstract}
\maketitle
\input{epsf.tex}
\section{Introduction}
The main motivation of this work is understanding  the mechanism of colour
confinement in QCD. An appealing possibility is dual superconductivity of the
ground state\cite{1,2,3}. ``Dual'' means that the role of electric and magnetic
quantities is interchanged with respect to ordinary superconductors: the
chromoelectric field acting between a $q \bar q$ pair is then channeled by dual
Meissner effect into Abrikosov flux tubes, which behave as
strings\cite{4,5,6}. Such tubes have been observed in lattice
configurations\cite{7}. The question is: can we directly detect dual
superconductivity on the lattice? The problem is always reduced to a $U(1)$
problem by the so called abelian projection\cite{1,8,9}: our task is then to
characterize dual superconductivity of a $U(1)$ system, and to find an
unambigous way to detect it.

Dual superconductivity is the spontaneous breaking \`a la Higgs of the $U(1)$
symmetry related to the conservation of magnetic charge: monopole charges
condense in the vacuum (in the same way as Cooper pairs do in ordinary
superconductors), making it $U(1)$ non invariant\cite{10}. If the vacuum is
$U(1)$ invariant the vacuum expectation value ({\it vev}) of any operator with
non zero magnetic charge is zero. Spontaneous breaking of $U(1)$ is therefore
signalled by non vanishing {\it vev} of any magnetically charged operator. Such
{\it vev} is called a disorder parameter in the language of statistical
mechanics\cite{11}. A rigorous proof exists\cite{12} that in compact $U(1)$ on
the lattice monopoles do condense for $\beta=1/e^2$ smaller than some critical
value $\beta_c$: the proof requires a specific form of the action (Villain) and
infinite volume, to perform the transformation to dual variables\cite{12bis}. The
difficulties with finite volume were already noticed in ref\cite{11} for the
Ising model. We have constructed a disorder parameter which coincides with that
of ref\cite{12} for Villain action and infinite volume, but can be used with
any form of the action and can detect dual superconductivity in finite lattices.
\section{Construction of the disorder parameter}
As magnetically charged operator we shall use a creation operator for
monopoles, defined as follows\cite{13}. Let $\Pi_i({\bf x},t) = F_{0i}({\bf x},t)$ be
the usual conjugate momentum to the field variables $A_i({\bf x},t)$. Then the
operator
\begin{equation}
\mu({\bf y},t) = {\rm exp}\left[{\rm i}\!\int\!{\rm d}^3{\bf x}\,
\Pi_i({\bf x},t)\frac{1}{e}b_i({\bf x}-{\bf y})\right]
\label{eq:1}
\end{equation}
creates a monopole if $b_i({\bf x}-{\bf y})$ is the vector potential produced
in ${\bf x}$ by a monopole of charge $m/e$ sitting in ${\bf y}$. Putting the
string along the 3 axis, and ${\bf r}$ $=$ ${\bf x}$ $-$ ${\bf y}$,
\begin{equation}
b_i({\bf x}-{\bf y}) = \frac{m}{2}\frac{\displaystyle
\varepsilon_{3ij} r_j}{\displaystyle r (r- r_3)}
\label{eq:2}\end{equation}
In fact $\mu$ is a translation operator of the field, as is easily seen in the
Schr\"odinger picture
\begin{equation}
\mu({\bf y},t)|A_i({\bf x},t)\rangle =
|A_i({\bf x},t)+ b_i({\bf x},{\bf y})\rangle
\label{eq:3}\end{equation}
In the same language if ${\bf \nabla}\wedge{\bf g} = 0$ the operator
\begin{equation}
\gamma(t) = {\rm exp}\left[\frac{{\rm i}}{e}\!\int\!{\rm d}^3{\bf x}\,
\Pi_i({\bf x},t)g_i({\bf x})\right]
\label{eq:4}
\end{equation}
defines a gauge transformation
\[\gamma(t)|A_i({\bf x},t)\rangle =
|A_i({\bf x},t)+ g_i({\bf x})\rangle\]
The correct prescription to define {\it vev} of $\mu$ in the euclidean region
with Feynman path integral can be checked on a system of free photons, with periodic
boundary conditions. It amounts to define the {\it vev} of the operator $\mu$ as\cite{13}
\begin{equation}
\hskip-15pt
\langle\bar\mu\rangle = \frac{\langle\displaystyle \mu({\bf y},y_4)\rangle}
{\langle\displaystyle \gamma(y_4)\rangle} =
\frac{\displaystyle
\int\!{\cal D} A_\mu{\rm exp}\left[-\beta (S+S_b)\right]}
{\displaystyle
\int\!{\cal D} A_\mu{\rm exp}\left[-\beta (S +S_g)\right]}
\label{eq:5}\end{equation}
where
\begin{eqnarray*} S_b &=&
\int\!{\rm d}^3{\bf x} F_{4i}({\bf x},y_4)  b_i({\bf x}-{\bf y})\\
S_g &=&
\int\!{\rm d}^3{\bf x} F_{4i}({\bf x})  g_i({\bf x})
\end{eqnarray*}
$S$ is the action and $g_i$ is any gauge configuration subjected to the
condition that
\[\int\!{\rm d}^4 k |\tilde{\bf b}(k)|^2 = \int\!{\rm d}^4 k |\tilde{\bf g}(k)|^2\]
The factor $\beta$ in the second term of the exponents come from the factor
$1/e$ in the magnetic charge of the monopole times the $1/e$ coming from the
usual rescaling of the fields, the same which produces the factor $\beta$ in
front of $S$. For free photons (or for $\beta > \beta_c$) $S$ $=$ $F_{\mu\nu}
F^{\mu\nu}/4$, the integral (\ref{eq:5}) is gaussian and
\begin{eqnarray}
\langle\bar\mu\rangle &=& {\rm exp}
\left[-\frac{\beta}{4}\int\!\!
\frac{\displaystyle {\rm d}^3 {\bf k}}
{\displaystyle (2\pi)^3}
|{\bf k}||\tilde{\bf b}
({\bf k})|^2
\right]\label{eq:7}\\
&=&C {\rm exp}\left[-\beta\ln(\frac{V}{a^3})\right]
\nonumber
\end{eqnarray}
$a$ is some ultraviolet cutoff. $\langle\bar\mu\rangle \to 0$ as $a\to 0$ or at
fixed $a$ (e.g. on a lattice) as $V\to \infty$ on the lattice. 
Our disorder parameter becomes
an order parameter of the dual theory, i.e. exactly zero for $\beta>\beta_c$,
only in the limit $V\to \infty$: with a finite number of degrees of freedom,
being analytic in $\beta$, it cannot be zero for $\beta > \beta_c$, since then
it would be identically zero\cite{11}. What is really important, however, 
in order to
detect superconductivity, is to show that it is different from zero below
$\beta_c$, independent of the finiteness of the volume. To accomplish this we
will compute $\langle\bar\mu\rangle$ not directly, but through
\begin{eqnarray}
\rho_b(\beta) &=& \frac{{\rm d}\ln \langle\bar\mu\rangle}{{\rm d}\beta}
 =\label{eq:8}\\
&=&\langle S + S_g\rangle_{S+S_g} - \langle S + S_b\rangle_{S+S_b}
\nonumber\end{eqnarray}
The label on the right of the brackets indicates the action used to compute the
average: eq.(\ref{eq:8}) directly follows from eq.(\ref{eq:5}). Since
$\langle\bar\mu\rangle_{\beta=0} = 1$, in terms of $\rho_b$ \begin{equation}
\langle\bar\mu\rangle = {\rm exp}\left(\int_0^\beta\!\rho_b(\beta'){\rm d}\beta'
\right)\label{eq:9}\end{equation}
\vskip0.2in
\noindent
\begin{minipage}{\linewidth}
\vbox to1.15\linewidth{
\epsfysize0.99\linewidth
{\centerline{
\epsfbox{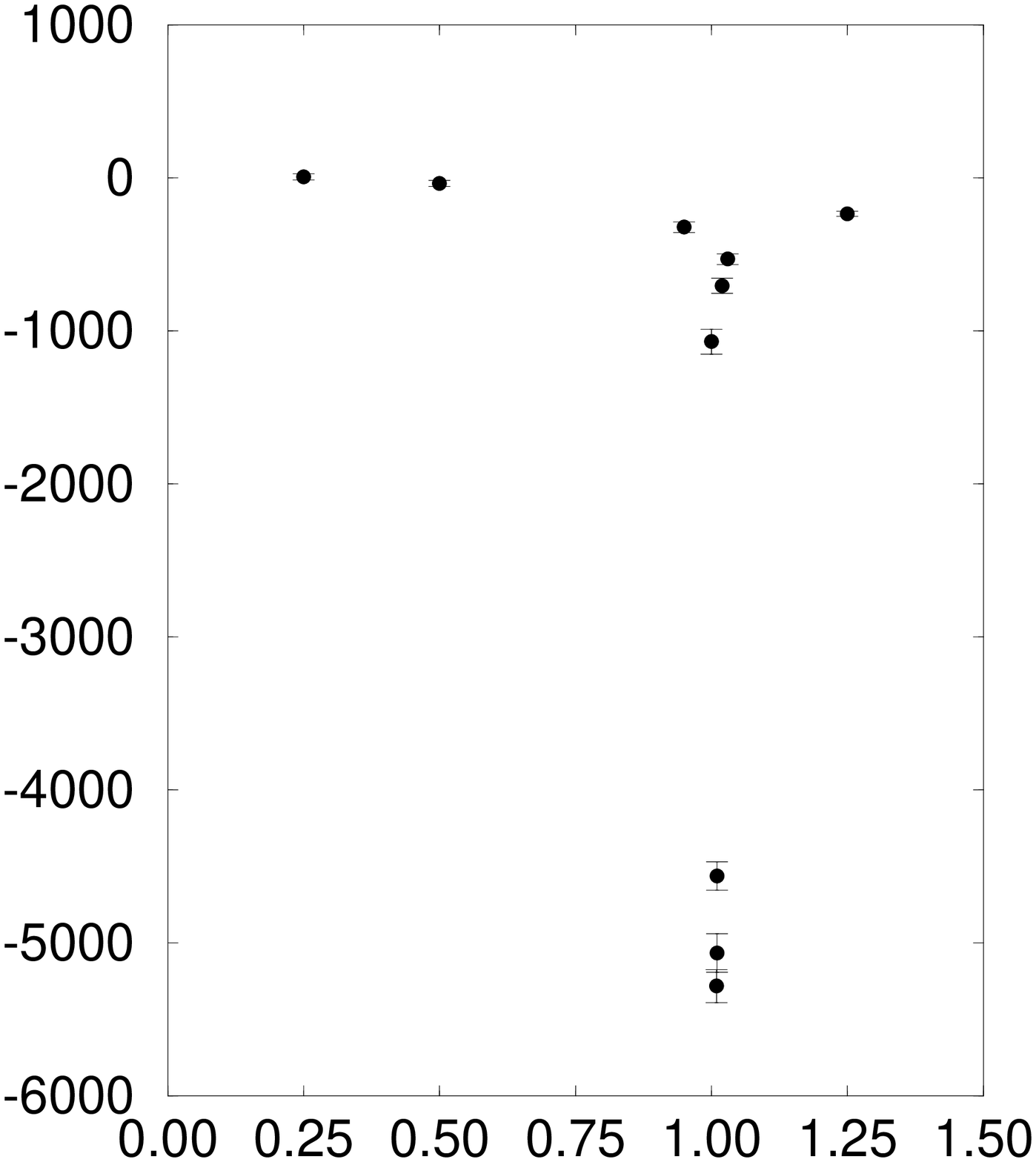}
}}
\vskip-0.5\linewidth
\hskip0.26\linewidth$\rho$
\vskip0.26\linewidth
\hskip0.72\linewidth$\beta$
\vskip0.15\linewidth
{\centerline
\noindent
{{\bf Fig.1:}$\rho_b(\beta)$ on a $12^4$ lattice. 
}}
}
\end{minipage}
\par\noindent
Computing $\langle\bar\mu\rangle$ from $\rho$ has a few advantages.
\begin{itemize}
\item[(i)] It solves the problem of 
fluctuations produced by the
fact that $\langle\bar\mu\rangle$, being the
average of the exponent of an extensive quantity, has a non gaussian
distribution\cite{14}.
\item[(ii)]
It
eliminates problems of boundary conditions in the definition of $\langle\bar\mu\rangle$\cite{12bis}. Indeed the r.h.s. of eq.(\ref{eq:8}) is the average of an extensive
quantity, and is sensitive to the bulk properties of the system
at least for $\beta < \beta_c$, i.e.
 in the presence of a mass gap.
Computing  $\langle\bar\mu\rangle$ with periodic and antiperiodic b.c. gives the same result
within errors.
\end{itemize}
To be rigorous we should have started from the correlation function betwen a
monopole - antimonopole pair, defining $\langle\bar\mu\rangle$ through the
cluster property
\[
\frac{\displaystyle
\langle\mu(x)\mu(y)\rangle}{\displaystyle\langle\gamma^2\rangle}
\mathop\simeq\limits_{|x-y|\to \infty}\langle \bar\mu\rangle^2
\]
Putting $S_{b\bar b} = S_b + S_{\bar b}$
\begin{eqnarray*}
\rho_{b\bar b}&=&\!
\frac{{\rm d}}{{\rm d}\beta}
\ln\frac{\displaystyle
\langle\mu(x)\mu(y)\rangle}{\displaystyle\langle\gamma^2\rangle}
=\\
&=&
2\langle S+S_g\rangle_{S+S_g} - \langle S + S_{b \bar b}\rangle_{S+S_{b \bar b}} - \langle
S\rangle_S\end{eqnarray*}
the cluster property reads then, for $|x-y|\to \infty$
\[\rho_{b\bar b} = 2 \rho_b\]
However if the correlation length is much smaller than the lattice size,
measuring $\rho_{b\bar b}(\beta)$ or $\rho_b(\beta)$ are equivalent procedures.

\vskip0.2in
\noindent
\begin{minipage}{\linewidth}
\vbox to1.15\linewidth{
\epsfysize0.99\linewidth
{\centerline{
\epsfbox{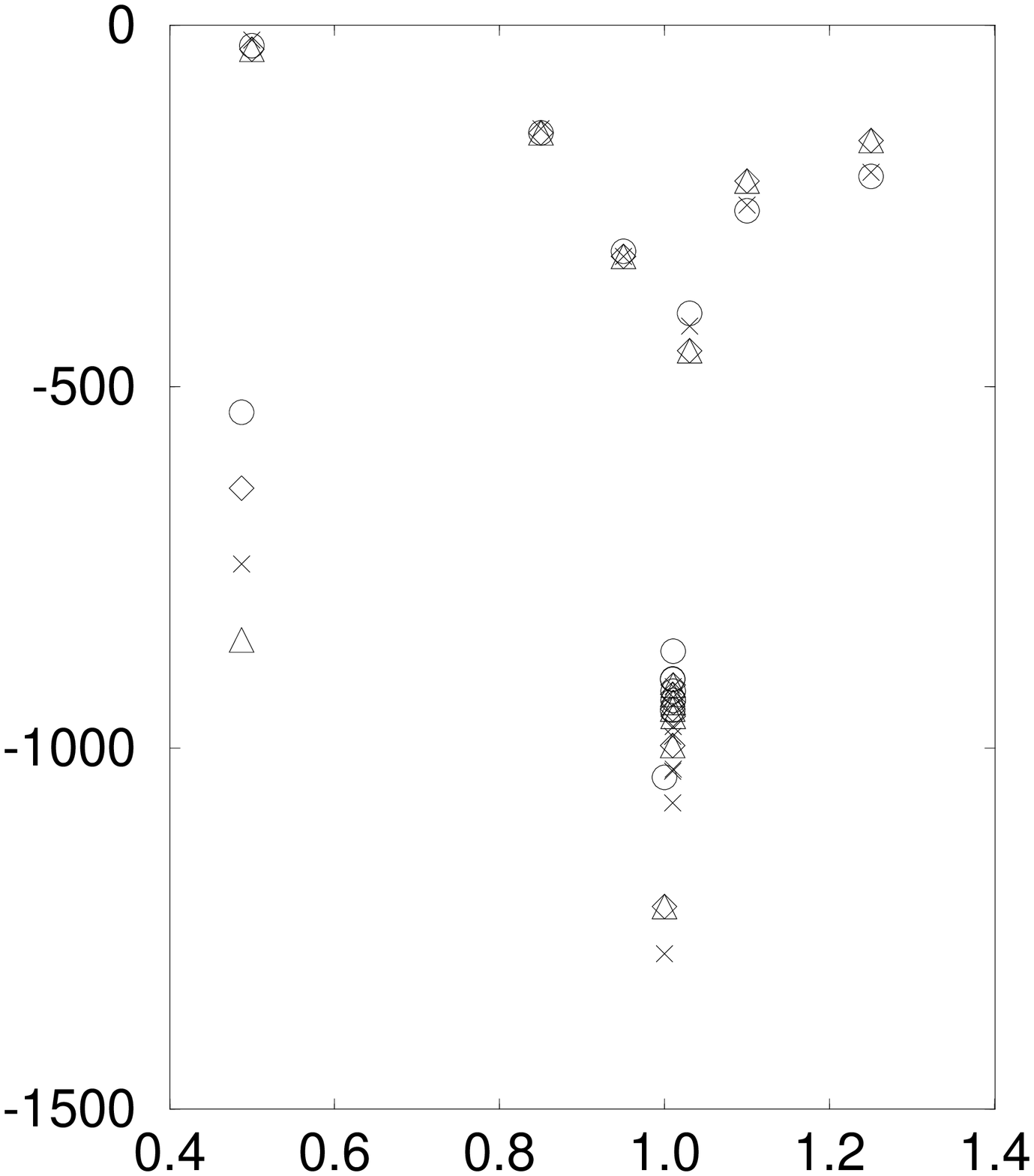}
}}
\vskip-0.66\linewidth
\hskip0.30\linewidth$\rho_{b\bar b}(4)$
\vskip0.1\linewidth
\vskip-0.1\linewidth
\hskip0.30\linewidth$\rho_{b\bar b}(7)$
\vskip0.1\linewidth
\vskip-0.1\linewidth
\hskip0.30\linewidth$\rho_{b\bar b}(12)$
\vskip0.1\linewidth
\vskip-0.1\linewidth
\hskip0.30\linewidth$2\rho_{b}$
\vskip0.8in\hskip0.75\linewidth$\beta$
\vskip0.3in
{\centerline
\noindent
{{\bf Fig.2:}$\rho_{b\bar b}(\beta)$ and $2\rho_b(\beta)$ on a
$6^3\times20$ latice. 
}}
}
\end{minipage}
\par\noindent

Numerical results on a $6^3\times20$ for $\rho_b$
and $\rho_{b\bar b}$ are shown in fig.2, and confirm the validity of the cluster property at
distances $d\geq 4$. $\rho_b$ for a lattice $12^4$ is shown in fig.1.

At large $\beta$'s $\rho$ agrees with eq.(\ref{eq:7}). A large negative peak at
$\beta = \beta_c$,
whose area is an increasing function of $V$, signals a vanishing $\langle\bar
\mu\rangle$. For $\beta<\beta_c$ $\langle\bar\mu\rangle$ becomes volume independent at
sufficiently large volumes, showing unambigously monopole condensation.

In conclusion we have a reliable disorder parameter which can be used with any
form of the action, at finite volume, to detect dual superconductivity of the
vacuum.


\begin{thebibliography}{99}
\bibitem{1} G. 't~Hooft, in ``High Energy Physics'', EPS
International Conference, Palermo 1975, ed. A.~Zichichi;
{\em Nucl Phys.} {\bf B 190} (1981) 455.
\bibitem{2}S. Mandelstam, {\em Phys. Rep.} {\bf 23C} (1976) 245.
\bibitem{3}G. Parisi, {\em Phys. Rev.} {\bf D 11} (1975) 970.
\bibitem{4}G. Veneziano, {\em Nuovo Cimento} {\bf 57 A} (1968) 190.
\bibitem{5}H.B. Nielsen, P. Olesen, {\em Nucl. Phys.} {\bf B61} (1973) 45.
\bibitem{6}Y. Nambu, {\em Phys. Rev.} {\bf D 10} (1974) 4262.
\bibitem{7}A. Di Giacomo, M.~Maggiore and \v{S}.~Olejn\'{\i}k,
{\em Phys. Lett.} {\bf B236} (1990) 199; {\em Nucl. Phys.} {\bf B347}
(1990) 441;
R.W. Haymaker, J. Wosiek, {\em Phys. Rev.} {\bf D 36}
(1987) 3297;
\bibitem{8}A.S. Kronfeld, G. Schierholz, U.J. Wiese, {\em Nucl.
Phys.} {\bf B293}(1987) 461.
\bibitem{9} L. Del Debbio, A. Di Giacomo, G. Paffuti, G. Pieri, {\em these
proceedings}.
\bibitem{10}S. Weinberg, {\em Progr. of Theor. Phys.
Suppl.\/} No. 86 (1986) 43.
\bibitem{11}L.P. Kadanoff, H.Ceva, {\em Phys. Rev.} {\bf B3} (1971) 3918.
\bibitem{12}J. Fr\"olich, P.A. Marchetti, {\em Comm. Math. Phys.}
{\bf 112} (1987) 343.
\bibitem{12bis}L. Polley, U. Wiese, {\em Nucl. Phys.} {\bf B 356}
(1991) 629.
\bibitem{13} L. Del Debbio, A. Di Giacomo, G. Paffuti, IFUP-TH- 16/94;
submitted for publication.
\bibitem{14} A. Hasenfratz, P. Hasenfratz, F. Niedermayer, {\bf Nucl. Phys.}
{\bf B 239} (1990) 739.
\end{thebibliography}
\end{document}